\documentclass[prl,a4paper, amsfonts, amssymb, amsmath, reprint, showkeys, nofootinbib, twoside,superscriptaddress]{revtex4-2}

\usepackage[english]{babel}

\usepackage{graphicx}
\usepackage{subcaption}
\usepackage{amsmath}
\usepackage{amsfonts}
\usepackage{soul}
\usepackage[utf8]{inputenc}
\usepackage{xcolor}
\usepackage{physics}
\usepackage{overpic,color}
\usepackage{hyperref}
\hypersetup{
    colorlinks=true,
    linkcolor=blue,
    filecolor=magenta,      
    urlcolor=blue,
    citecolor = blue,
    pdftitle={Molecular symmetry controls photoelectron time delay and hidden phase topology},
    pdfpagemode=FullScreen,
    }
\usepackage{dsfont}
\usepackage{mathtools}
\usepackage{bm}
\usepackage[normalem]{ulem}
\usepackage{lipsum}

\DeclareMathAlphabet{\pazocal}{OMS}{zplm}{m}{n}
\usepackage{isomath}

\renewcommand{\BibitemShut}[1]{}

\usepackage[scaled=0.96,lining]{XCharter}
\usepackage{newtxmath}

\usepackage{caption}
\usepackage{ragged2e}

\DeclareCaptionJustification{fulljustified}{\justifying}

\begin{document}
\title{Universal Threshold Memory in Bound-State Recapture}
\author{Sajad Azizi}
\email{azizi.dajas@gmail.com}
\affiliation{Leibniz Supercomputing Centre, Boltzmannstra{\ss}e 1, 85748 Garching bei M{\"u}nchen, Germany}

\date{\today}

\begin{abstract}
Driven slowly, a quantum system forgets how it was driven. We show that a bound state taken slowly to its continuum threshold and back does not. The recaptured fraction depends on the ratio of the approach and return rates, however slow both are. The two rates set the sizes of two threshold packets whose phase-sensitive overlap fixes recapture. Common slowing preserves their size ratio. A factorized treatment of approach and return misses this dependence. For a short-range $s$-wave state whose inverse scattering length touches zero quadratically in time, we obtain the dependence in closed form, with the maximum $1/\varphi^2$ at matched rates, where $\varphi$ is the golden ratio. A ratio of three recaptured fractions makes this threshold memory accessible with finite pulses. For radio-frequency-dressed $^{85}$Rb halo dimers with half-pulse widths of 30 and $120~\mu$s, coupled-channel calculations give this ratio within $0.5\%$ of its slow-limit value.
\end{abstract}

\maketitle

\emph{Introduction.}---Slow driving is a standard method to suppress transitions between instantaneous eigenstates. When a state is separated from the rest of the spectrum by a gap, it follows the changing Hamiltonian~\cite{BornFock1928}, and the probability of leaving it vanishes as the drive slows—exponentially so for a Landau–Zener passage through an avoided crossing~\cite{LandauZener1932}. At a critical point where the gap closes, the density of excitations still decreases as a power of the rate~\cite{KibbleZurek,ZurekDornerZoller2005}. However, at a continuum threshold, the situation is different. For a short-range $s$-wave state, the size of the instantaneous bound state diverges as its energy reaches the continuum, without a neighboring discrete level to cross. When the binding is restored, only part of the state is recaptured, no matter how slowly the passage is made. This loss has been recognized since early studies of electron detachment in slow collisions~\cite{Demkov1964,DemkovOstrovskii1988}. If the inverse scattering length touches zero quadratically in time, with equal rates of approach and return, the recapture probability is known exactly: $1/\varphi^2\simeq0.382$, where $\varphi$ is the golden ratio~\cite{DevdarianiDemkov1971,Devdariani1972,SokolovskiPons2015,SokolovskiPons2016}.

We ask what the recaptured fraction remembers of the passage as the absolute rates vanish. For a touching with different powers of $t$ on approach and return, Ref.~\cite{SokolovskiPons2016} proposed a factorized expression, with one factor per half and no remaining rate dependence. The rates still matter when the touching is quadratic on both sides: the recaptured fraction depends on their ratio. We derive this dependence in closed form for an isolated short-range $s$-wave threshold, where all short-range interactions share the same slow limit. The recaptured fraction has its maximum, $1/\varphi^2$, at equal rates and is unchanged when both rates are reduced together. It tends to zero when either rate is taken to zero with the other held fixed. We call this surviving dependence \emph{threshold memory}.

\begin{figure*}[!t]
\includegraphics[width=\textwidth]{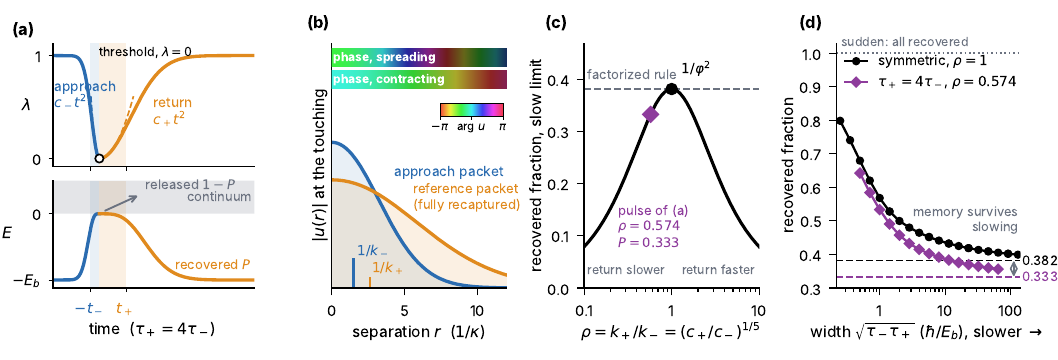}
\captionsetup{justification=fulljustified,singlelinecheck=false}
\caption{Threshold memory, for a bound state of initial inverse scattering length $\kappa$ and binding energy $E_b=\hbar^2\kappa^2/m$. (a) An inverse scattering length $\lambda(t)$, in units of $\kappa$, that touches zero once, with quadratic coefficients $c_-$ on approach and $c_+$ on return (dashed; here $\tau_+=4\tau_-$, $c_+=c_-/16$), and the bound level $E=-\hbar^2\lambda^2/m$ meeting the continuum; shaded, the windows $|t|<t_\pm$ in which the state can no longer follow. A fraction $P$ is recaptured, $1-P$ released. (b) For the pulse of (a) with $\tau_-=2\hbar/E_b$ [$\lambda$ as in (d)], the packet prepared by the approach (size $1/k_-$) and the packet the return recaptures completely (size $1/k_+$), $k_\pm\propto c_\pm^{1/5}$, phases as color bands (hue $=\arg u$); one spreads, the other contracts. (c) The law, Eq.~\eqref{eq:law}, against $\rho=k_+/k_-$; circle, equal rates; diamond, the pulse of (a); dashed, the factorization of Ref.~\cite{SokolovskiPons2016} extended to unequal rates. (d) Recapture from Eq.~\eqref{eq:robin} with $\lambda=\kappa[1-e^{-t^2/2\tau_\pm^2}]$ for the symmetric pulse and the pulse of (a), against the width $\sqrt{\tau_-\tau_+}$ in units of $\hbar/E_b$~\cite{SM}; each tends to its own floor, the markers of (c).}
\label{fig:story}
\end{figure*}

The memory comes from a competition of scales, Fig.~\ref{fig:story}. A bound state with inverse scattering length $\lambda$ has size $1/\lambda$ and binding energy $\hbar^2\lambda^2/m$. Adiabatic following breaks down when this energy becomes comparable to $\hbar/|t|$, the energy resolution of the time $|t|$ remaining before the touching. For $\lambda=ct^2$, this Kibble--Zurek freeze-out estimate~\cite{KibbleZurek} gives the time scale $t_*\propto c^{-2/5}$ and the characteristic size $1/k_*\propto c^{-1/5}$ of the spreading packet at the touching. The return recaptures completely only the time reverse of the threshold packet prepared by an approach at the return rate. The recapture amplitude is the overlap between this selected packet and the packet produced on approach. Since the threshold has no intrinsic length, slowing both halves enlarges both packets by the same factor, and only their size ratio survives.

To test the law we need an isolated threshold touching and a readout that tolerates pulses far from the slow limit. Radio-frequency dressing of a halo dimer~\cite{Kohler2006,Chin2010,Kaufman2009,Hanna2010,Tscherbul2010,Papoular2010} provides the first. The pulse mixes the dimer's spin channel with two others that remain energetically closed throughout, bringing the binding to zero while the dressed spin gaps stay open. Its rise and fall set the two rates at a fixed magnetic field. For the readout we use a ratio of three recaptured fractions. It equals one when approach and return contribute independent factors with the same rate dependence. For the pulses studied, our calculations show that the ratio removes most of the deviation from the slow limit, allowing the threshold law to be tested with finite pulses.

\emph{Two scales at one threshold.}---Near an isolated short-range $s$-wave threshold, the low-energy relative motion of two atoms is described by the zero-range model
\begin{equation}
\mathrm{i}\hbar\partial_tu=-\frac{\hbar^2}{m}\partial_r^2u,\qquad
\partial_ru(0,t)=-\lambda(t)u(0,t),
\label{eq:robin}
\end{equation}
where $r>0$ is the interatomic separation, $m$ the atomic mass, $u(r,t)/r$ the radial wave function, and $\lambda(t)=1/a(t)$ the inverse scattering length. The interaction enters only through the boundary condition. For $\lambda>0$, the instantaneous bound state $u_b\propto e^{-\lambda r}$ has energy $E=-\hbar^2\lambda^2/m$ and reaches the continuum at $\lambda=0$. We place the touching at $t=0$ and take $\lambda=c_-t^2$ on approach ($t<0$) and $\lambda=c_+t^2$ on return ($t>0$). Here $c_\pm>0$ are the approach and return rates, and the binding energy vanishes as $t^4$. If $\lambda(t)$ is twice differentiable at the touching and has nonzero curvature, its leading quadratic coefficient is the same on both sides; other touching powers are treated in Appendix A.

Each half sets a length $1/k_\pm$ and a time $t_\pm$, with $k_\pm=(m^2c_\pm/\hbar^2)^{1/5}$ and $t_\pm=m/(\hbar k_\pm^2)$. In these units, $\lambda/k_\pm=(t/t_\pm)^2$, and the rate drops out of Eq.~\eqref{eq:robin}. Unequal coefficients give a jump in $\ddot\lambda$ at $t=0$; smoothing the junction over a time short compared with both $t_-$ and $t_+$ gives negligible corrections~\cite{SM}. As the rates vanish, $k_\pm r_0\to0$ for any finite interaction range or effective range $r_0$. The zero-range model therefore becomes exact at an isolated $s$-wave threshold in the slow limit, and the law below is independent of the short-range interaction. An approach starting in the bound state well before the touching produces $u_-(r)=\sqrt{k_-}\,\Phi(k_-r)$ at $t=0$, where $\Phi$ is the universal packet of the scaled problem, normalized to $\int_0^\infty|\Phi(X)|^2\,\mathrm{d}X=1$. The return, run backward in time, is an approach with rate $c_+$. Because the Hamiltonian is real, the state it recaptures completely is the time reverse of its own threshold packet, $u_{\rm ref}(r)=\sqrt{k_+}\,\Phi^*(k_+r)$, up to a phase. The recapture amplitude is their overlap~\cite{Demkov1964,SokolovskiPons2016},
\begin{equation}
A=\int_0^\infty u_{\rm ref}^*(r)\,u_-(r)\,\mathrm{d}r
 =\sqrt{\rho}\int_0^\infty\Phi(X)\,\Phi(\rho X)\,\mathrm{d}X,
\end{equation}
with $X=k_-r$ and $\rho=k_+/k_-=(c_+/c_-)^{1/5}$. The conjugation in the inner product cancels that from time reversal, making the overlap bilinear in $\Phi$. Even at $\rho=1$, the integrand is $\Phi^2$ rather than $|\Phi|^2$: the phase variation across the packet prevents complete recapture when the sizes match.

Evaluating the overlap with the threshold packet selected under the regularity assumption in Appendix A gives
\begin{equation}
P(\rho)=|A|^2=
\frac{4\rho}{5}
\left[
\frac{1+\varphi\rho+\rho^2}
{(1+\rho)(1+\rho/\varphi+\rho^2)}
\right]^2,
\label{eq:law}
\end{equation}
where $\varphi=(1+\sqrt5)/2$ is the golden ratio. Matched rates give the unique maximum, $P(1)=1/\varphi^2\simeq0.382$. The law is symmetric under exchange of the rates, $P(\rho)=P(1/\rho)$, and gives $P\simeq4\rho/5$ for $\rho\ll1$. Equation~\eqref{eq:law} has no adjustable parameter. Direct propagation of Eq.~\eqref{eq:robin} from the initial bound state and an independent evaluation of the overlap agree with this law to within $2\times10^{-6}$ in absolute probability at the tabulated values of $\rho$~\cite{SM}. The same construction extends to $\lambda=c_\pm|t|^\nu$ with positive integer $\nu$ (Appendix A), where $\rho=(c_+/c_-)^{1/(2\nu+1)}$, and again gives the maximum at matched rates. A linear touching requires a kink in the drive at the threshold and raises this maximum to $16/27$.
The asymmetric passages of Ref.~\cite{SokolovskiPons2016} combined different touching powers with a single rate parameter. Extending their rate-independent factors (Eq.~(D3) there) to quadratic touching with unequal rates would give $P=1/\varphi^2$ for every $\rho$ (dashed curve in Fig.~\ref{fig:story}(c)). The full overlap keeps the dependence on the rate ratio that this factorization drops.

\begin{figure*}[t]
\includegraphics[width=\textwidth]{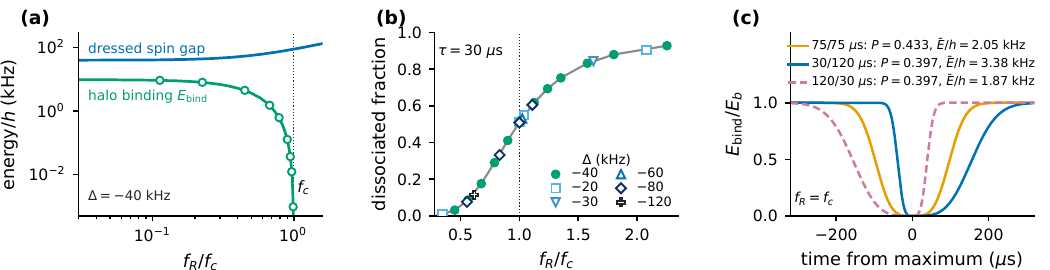}
\captionsetup{justification=fulljustified,singlelinecheck=false}
\caption{Realization with RF-dressed $^{85}$Rb dimers. (a) Binding energy $E_{\rm bind}$ of the dressed halo state against the drive $f_R/f_c$ at $\Delta=-40$~kHz: it vanishes at $f_c$, while the dressed spin gap, the spacing of the dressed spin thresholds, keeps growing. Line: three-channel contact model with effective ranges, scaled by its own $f_c$; circles: stationary coupled channels. (b) Fraction dissociated by $30$-$\mu$s Gaussian pulses against the peak drive $f_R/f_c$. Symbols: coupled channels, a fourteen-pulse scan at $-40$~kHz (filled circles, joined by the line) and eleven pulses at five further detunings from $-20$ to $-120$~kHz. (c) Binding energy, from the contact model of (a), during split-Gaussian pulses peaking at $f_c$, $\Delta=-40$~kHz, labeled by the widths $\tau_-/\tau_+$ in $\mu$s, the coupled-channel survival $P$, and the mean energy $\bar E$ of the released pairs.}
\label{fig:halo}
\end{figure*}

\emph{Reading the memory with finite pulses.}---Lengthening the pulses brings the system toward the slow limit only gradually, $t_\pm/\tau_\pm\propto\tau_\pm^{-1/5}$. For the halo dimers considered below, convergence toward this limit remains slow on the scale of their measured undriven lifetime. We therefore read the memory through a ratio of three survivals. The survival $P(\tau_-,\tau_+)$ is the probability of recapturing the initial bound state after a pulse with approach width $\tau_-$ and return width $\tau_+$. The two halves have the same shape, stretched to their respective widths, and the same peak amplitude, at which the binding vanishes. Using two symmetric reference pulses with half-pulse widths $\tau_-$ and $\tau_+$ at this peak amplitude, we define the normalized survival
\begin{equation}
M=
\frac{P(\tau_-,\tau_+)}
{\sqrt{P(\tau_-,\tau_-)\,P(\tau_+,\tau_+)}}
\;\longrightarrow\;
M_\infty=\frac{P(\rho)}{P(1)},
\label{eq:M}
\end{equation}
where the arrow denotes the slow limit $\tau_\pm\gg t_\pm$ at fixed width ratio. For these pulse shapes, $c_\pm\propto\tau_\pm^{-2}$, so $\rho=(\tau_-/\tau_+)^{2/5}$.

The normalization tests the factorized description directly. If each half contributed an independent factor, $P(\tau_-,\tau_+)=F(\tau_-)F(\tau_+)$ with the same $F$ on approach and return, then $M=1$ identically, whereas the law gives $M_\infty<1$ for unequal rates. Multiplicative loss factors of this same form cancel in $M$, as does a detection efficiency common to the three measurements. Most of the finite-pulse correction also cancels for the pulses studied. For Gaussian halves, the first correction to the quadratic touching is quartic in $t$, with relative size set by $e_\pm=(t_\pm/\tau_\pm)^2$ on the threshold time scale. To first order in $e_\pm$, the return and approach contribute relative corrections $\mathcal C(\rho)\,e_+$ and $\mathcal C(1/\rho)\,e_-$, respectively. For the symmetric references both coefficients are $\mathcal C(1)$, so the denominator of $M$ removes $\mathcal C(1)(e_-+e_+)$. Only the variation of $\mathcal C$ with $\rho$ remains at this order. Near equal rates, $|\ln\rho|\ll1$, this gives
\begin{equation}
\frac{M}{M_\infty}-1
\simeq(A_1+2A_2)\,e_G\ln^2\rho,
\label{eq:Mfirst}
\end{equation}
where $G=\sqrt{\tau_-\tau_+}$ is the geometric width, $t_G$ its threshold time, and $e_G=(t_G/G)^2$. With $x=\ln\rho$, $A_1$ and $A_2$ are the coefficients of $x$ and $x^2$ in the Taylor expansion of $\mathcal C(e^x)$ about $x=0$~\cite{SM}. The leading correction falls as $G^{-2/5}$ at fixed width ratio. The full relative deviation is exactly even in $\ln\rho$ at fixed $G$, because reversing the pulse does not change the survival.

\emph{Realization with RF-dressed halo dimers.}---We take a halo dimer of two $^{85}$Rb atoms in the state $a=|f=2,m_f=-2\rangle$ (zero-field labels) at $B=156.840$~G, near the broad 155~G Feshbach resonance~\cite{Claussen2003,Blackley2013}. Its binding energy is $E_b/h=9.770$~kHz and its size $1/\kappa=2.1\times10^3\,a_0$, with $E_b=\hbar^2\kappa^2/m$. A linearly polarized RF field perpendicular to the static field couples $a$ to $b=|2,-1\rangle$, which lies $78.37$~MHz below it. We choose the RF frequency above the atomic transition, $\Delta=\nu_{\rm at}-\nu_{\rm RF}<0$. In the rotating frame, the pair thresholds $ab$ and $bb$ then lie above $aa$ by $-h\Delta$ and $-2h\Delta$, and both channels stay closed. The drive mixes the halo channel with these two closed channels. We write its strength as $f_R=g_S\mu_BB_{\rm pk}/2h$, where $B_{\rm pk}$ is the peak RF amplitude, $g_S$ the electron $g$ factor, and $\mu_B$ the Bohr magneton. The closed-channel scattering lengths are negative, $a_{ab}=-557\,a_0$ and $a_{bb}=-664\,a_0$, whereas $a_{aa}=2194\,a_0$~\cite{SM}. The mixing reduces the binding energy relative to the lowest dressed threshold until it vanishes at a critical drive $f_c$, where the scattering length $a_{\rm eff}$ in that channel diverges [Fig.~\ref{fig:halo}(a)]. The spacing of the dressed spin thresholds, $h\sqrt{\Delta^2+d^2f_R^2}$ with $d=0.357$ the dimensionless atomic matrix element, increases with the drive. The binding gap closes and the spin gap stays open, so a pulse can be slow for the spins and still nonadiabatic for the relative motion at the threshold.

Coupled-channel calculations with the potentials of Ref.~\cite{Strauss2010} give $f_c=222.0$~kHz at $\Delta=-40$~kHz, corresponding to a peak RF field of $0.16$~G. A three-channel zero-range model, which matches the contact conditions of the bare channels to decaying solutions in the dressed channels~\cite{Moerdijk1996,Lavoine2021,Tononi2026}, gives $f_c=213.8$~kHz. We use this model below with the pulse peak at its own threshold~\cite{SM}. After scaling the drive by $f_c$, the trajectory depends only weakly on detuning. For $|\Delta|=20$--$160$~kHz and $1/2\le f_R/f_c\le1$, the sampled coupled-channel curves of $a_{aa}/a_{\rm eff}$ lie within $0.03$ of the common approximation described in the SM~\cite{SM}. The calculated dissociation probabilities for eleven $30$-$\mu$s pulses at five other detunings between $-20$ and $-120$~kHz lie within $0.01$ of the $-40$~kHz curve, see Fig.~\ref{fig:halo}(b).

A Gaussian envelope $\chi(t)=e^{-t^2/(2\tau^2)}$ with peak drive $f_c$ gives the quadratic touching. Near the peak, $f_c\chi(t)\simeq f_c[1-t^2/(2\tau^2)]$. If the spins follow the envelope ($2\pi|\Delta|\tau\gtrsim3$), the inverse scattering length follows the static curve: $\lambda(t)=1/a_{\rm eff}\simeq s\,t^2/(2a_{aa}\tau^2)$. In the three-channel model, $a_{aa}=1/\kappa$ matches the initial binding energy, and $s=1.18$ is minus the slope of $a_{aa}/a_{\rm eff}$ against $f_R/f_c$ at the threshold~\cite{SM}. Two Gaussian halves of widths $\tau_-$ and $\tau_+$ join at the peak with continuous $\chi$ and $\dot\chi$, giving $c_\pm=s/(2a_{aa}\tau_\pm^2)$ and hence $\rho=(c_+/c_-)^{1/5}=(\tau_-/\tau_+)^{2/5}$. The slope cancels, so the width ratio alone sets the scale ratio in Eq.~\eqref{eq:law}.

\begin{figure}[t]
\includegraphics[width=\columnwidth]{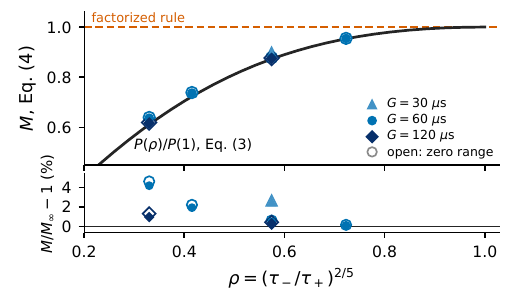}
\captionsetup{justification=fulljustified,singlelinecheck=false}
\caption{Normalized survival of finite Gaussian pulses against the scale ratio $\rho=(\tau_-/\tau_+)^{2/5}$, at $\Delta=-40$~kHz. Filled symbols: coupled channels; open symbols: three-channel zero-range model; triangles, circles, and diamonds: geometric widths $G=\sqrt{\tau_-\tau_+}$ of $30$, $60$, and $120~\mu$s. Upper panel: $M$, Eq.~\eqref{eq:M}; solid curve: the slow-limit law $M_\infty=P(\rho)/P(1)$; dashed line: $M=1$, the factorized rule. Lower panel: $M/M_\infty-1$ in percent.}
\label{fig:M}
\end{figure}

\emph{Coupled-channel test.}---We propagate the pulses at $\Delta=-40$~kHz with finite-range coupled-channel potentials based on Ref.~\cite{Strauss2010}. We calibrate them using only the initial binding energy. For approach and return widths of $30$ and $120~\mu$s, respectively, the survival is $0.3974$. The symmetric $30/30$ and $120/120~\mu$s reference pulses give $0.4896$ and $0.4196$, with a geometric mean of approximately $0.4532$. The normalized survival is therefore $M=0.8769$, $0.5\%$ above $M_\infty=0.8726$ at $\rho=0.574$, while the survival itself is $19\%$ above its limit $P(\rho)=0.3333$ (Fig.~\ref{fig:M}). Because the coupled-channel calculation retains finite-range interactions and spin dynamics at a dressed threshold, the agreement in $M$ tests the universal prediction beyond the zero-range model of Eq.~\eqref{eq:robin}.

The threshold times, $t_-\simeq33~\mu$s and $t_+\simeq99~\mu$s, are comparable to the pulse widths. Lengthening the pulses brings only slow improvement: the symmetric survival drops from $0.490$ at $30~\mu$s to $0.406$ at $240~\mu$s, still well above $1/\varphi^2$. Even a Gaussian half-pulse width as long as the measured undriven lifetime, $10.4(1.7)$~ms at $156.6$~G~\cite{ThompsonDecay}, would leave $t_\pm/\tau_\pm\simeq0.34$. For the pulses studied, the deviation of $M$ from $M_\infty$ grows with the width ratio at fixed $G$ and decreases with $G$ at fixed ratio, consistent with the trend in Eq.~\eqref{eq:Mfirst}. Because $e_G\simeq0.7$--$0.9$ at $G=60$--$120~\mu$s, we determine the deviations numerically~\cite{SM}. At $G=60~\mu$s, the deviation increases from $0.13\%$ for $\tau_+/\tau_-=2.25$ to $4.2\%$ for $16$. At $G=120~\mu$s, it is $0.34\%$ and $0.98\%$ for ratios $4$ and $16$, respectively (Fig.~\ref{fig:M}).

The three-channel model reproduces the coupled-channel $M$ within $0.42\%$ for every pulse tested. The static trajectory $\lambda(t)=1/a_{\rm eff}[f_c\chi(t)]$, used in the single-channel Eq.~\eqref{eq:robin}, reproduces $M$ within $0.54\%$~\cite{SM}. For the $30/120~\mu$s pulse, we use this single-channel model to test parameter sensitivity, keeping the widths fixed and resetting the peak to the recalculated $f_c$ for each parameter set. Varying $E_b/h$ by $\pm0.35$~kHz, or sampling the ranges of $a_{ab}$ and $a_{bb}$ spanned by seven interaction variants, changes $M$ by less than $5\times10^{-4}$. Combining the estimated parameter, model, and numerical contributions in quadrature gives $M=0.8769\pm0.0013$ for this pulse~\cite{SM}, excluding losses, interactions with residual atoms, and calibration errors.

One control distinguishes the predicted recapture from loss governed solely by the instantaneous amplitude; a second tests symmetry under pulse reversal. The $75/75$ and $30/120~\mu$s pulses have the same peak and the same sum of half-widths, $150~\mu$s, so with Gaussian halves they spend the same total time in every amplitude interval. A loss rate depending only on the instantaneous amplitude therefore predicts equal survival, whereas the coupled-channel calculations give $0.4334$ and $0.3974$, a difference of $3.6$ percentage points, see Fig.~\ref{fig:halo}(c). For the second control, reciprocity requires the $30/120$ and $120/30~\mu$s pulses to give the same survival, even with such a loss rate; the calculated survivals differ by less than $2\times10^{-13}$. A measured difference could arise from unequal pulse distortions, losses that depend on the passage, or differences in preparation or detection. The released pairs nevertheless carry different mean energies, $\bar E/h=3.38$ and $1.87$~kHz, respectively. Peak-amplitude errors and other envelope shapes are examined in the SM~\cite{SM}.

\emph{Experimental test.}---The proposed test requires the three survivals in Eq.~\eqref{eq:M}, measured after the asymmetric pulse and the two symmetric references at the same peak drive and a fixed magnetic field, for several width ratios. Halo dimers of $^{85}$Rb~\cite{ThompsonAssociation,Hodby2005} and Gaussian RF pulses on this transition~\cite{Wild2012} have been demonstrated separately; RF fields of the required strength have been achieved with $^6$Li~\cite{Scazza2025}. The threshold drive can be calibrated by fitting an amplitude scan of the recaptured signal to the calculated coupled-channel response. For readout, surviving dimers would be converted into atoms before imaging. A matched reference sequence would include a hold far from resonance before conversion, so that the dimers decay into atoms that escape detection while the unbound atoms remain. The difference between the two atom counts is proportional to the surviving dimer number. The calibration, detection, and loss budgets are given in the SM~\cite{SM}.

\emph{Conclusion.}---For a quadratic touching of the inverse scattering length at an isolated short-range $s$-wave threshold, the slow-limit recapture probability depends only on the ratio of the approach and return rates. Equation~\eqref{eq:law} gives this dependence in closed form under the regularity assumption in Appendix A, independently of the short-range interaction. The ratio of three survivals provides a finite-pulse test of this limit: for RF-dressed $^{85}$Rb dimers, coupled-channel calculations give a normalized survival within $0.5\%$ of its slow-limit value for $30/120~\mu$s pulses. A magnetic-field implementation is outlined in the SM~\cite{SM}. We conjecture that thresholds without an intrinsic scale retain analogous memory, with their own exponent and scaling function. Thresholds behind a centrifugal or Coulomb barrier, where the state stays normalizable, are outside this argument. Pulses with a linear cusp at their maximum, such as triangular or two-sided exponential envelopes, give a linear touching. The numerically tested extension in Appendix A then predicts a matched-rate limit of $16/27$, with $\rho=(c_+/c_-)^{1/3}$.

\makeatletter\let\auto@bib@innerbib\@empty\makeatother

\section*{End Matter}

\emph{Appendix A: Threshold packet and recapture.}---The recapture probability for a symmetric passage is established in Refs.~\cite{DevdarianiDemkov1971,Devdariani1972,DemkovOstrovskii1988,SokolovskiPons2015}. To describe different approach and return rates we need the wave packet at the touching, including its spatial phase. We construct it for the canonical quadratic approach, evaluate its overlap with a rescaled copy, and extend the construction to other touching powers.

\emph{Boundary value during the approach.}---We use canonical units, $\hbar=m=1$ and $c_-=1$, let $\vartheta=-t\ge0$ be the time remaining before the touching, and write $g(\vartheta)=u(0,-\vartheta)$ for the boundary value, so that $\vartheta=0$ is the touching and large $\vartheta$ the early part of the approach. Free propagation on the half-line combined with the contact condition gives the integral equation
$$
g(\vartheta)=\frac{e^{\mathrm{i}\pi/4}}{\sqrt\pi}\int_\vartheta^\infty
\frac{\sigma^2 g(\sigma)}{\sqrt{\sigma-\vartheta}}\,\mathrm{d}\sigma ,
$$
and starting in the adiabatic bound state requires
$$
g(\vartheta)\sim\sqrt2\,\vartheta\,e^{-\mathrm{i}\vartheta^5/5},\qquad \vartheta\to\infty,
$$
up to a constant phase. The Mellin transform $\mathcal G(w)=\int_0^\infty\vartheta^{w-1}g(\vartheta)\,\mathrm{d}\vartheta$, with $w$ complex, turns the integral equation into a recurrence; with the oscillatory integrals understood as regularized limits,
$$
\mathcal G(w)=e^{\mathrm{i}\pi/4}\,\frac{\Gamma(w)}{\Gamma(w+\tfrac12)}\,\mathcal G(w+\tfrac52),
$$
where $\Gamma$ is the Gamma function. A particular solution for the quadratic touching is
$$
\begin{gathered}
\mathcal G_2(w)=\beta^w Q(w),\\
Q(w)=\frac{\Gamma(w)}{\Gamma[(2w+1)/5]\,\Gamma[(2w+3)/5]},\\
\beta=\left(\frac25\right)^{4/5}e^{-\mathrm{i}\pi/10}.
\end{gathered}
$$
The recurrence does not fix the solution, since multiplication by a periodic function leaves it unchanged,
$$
\mathcal G(w)=C\,\varpi(w)\,\mathcal G_2(w),\qquad \varpi(w+\tfrac52)=\varpi(w),
$$
with $C$ a constant; the remaining step is to determine $\varpi$.

\emph{Selecting the solution.}---We assume that the early-time boundary value has the form
$$
g(\vartheta)=\vartheta\left[a_-(\vartheta)e^{-\mathrm{i}\vartheta^5/5}+a_+(\vartheta)e^{\mathrm{i}\vartheta^5/5}\right],\qquad \vartheta\gg1,
$$
with bounded amplitudes $a_\pm(\vartheta)$ of bounded variation, that is, of finite total variation over the large-$\vartheta$ tail. Within this class, bounds on the growth of $\mathcal G(w)$ force $\varpi$ to be constant~\cite{SM}, and the boundary value is proportional to $g_2$, the inverse Mellin transform of $\mathcal G_2$, a single Meijer $G$ function whose normalization is fixed by the initial bound state. The argument selects a unique solution within the stated class; that the physical solution belongs to the class has not been proved. Direct propagation from the adiabatic bound state tests the selected packet and its recapture law independently~\cite{SM}.

\emph{From the boundary value to recapture.}---Once $g$ is known, free propagation gives the packet at the touching,
$$
\Phi(X)\propto\int_0^\infty\vartheta^{3/2}e^{\mathrm{i}X^2/(4\vartheta)}g(\vartheta)\,\mathrm{d}\vartheta,
$$
with $X$ the separation in canonical length units and the constant chosen so that $\int_0^\infty|\Phi(X)|^2\,\mathrm{d}X=1$. Its spatial Mellin transform, $U(z)=\int_0^\infty X^{z-1}\Phi(X)\,\mathrm{d}X$, follows from the propagator expression,
$$
U(z)\propto2^z e^{\mathrm{i}\pi z/4}\,\Gamma(z/2)\,\mathcal G(\tfrac52+\tfrac z2).
$$
The recapture amplitude is the bilinear overlap of the two scaled packets, which the Mellin--Parseval identity writes as
$$
A(\rho)=\frac{\sqrt\rho}{2\pi\mathrm{i}}\int_{\gamma-\mathrm{i}\infty}^{\gamma+\mathrm{i}\infty}
U(z)\,U(1-z)\,\rho^{-z}\,\mathrm{d}z,\qquad 0<\gamma<1,
$$
along a vertical contour, with no complex conjugation in the integrand, as the amplitude requires. With the selected solution and $y=z/2$ the Gamma functions reduce to
$$
\begin{aligned}
&\Gamma(y)\,\Gamma(\tfrac12-y)\,Q(\tfrac52+y)\,Q(3-y)\\
&\qquad=\frac{625}{8}\,\frac{\sin[\pi(6+2y)/5]\,\sin[\pi(8+2y)/5]}{\sin(2\pi y)}
\end{aligned}
$$
on $0<\mathrm{Re}\,y<1/2$. Apart from the prefactor $\sqrt{\rho}$ and the factor $\rho^{-2y}$, all omitted factors are independent of $y$ and $\rho$. Summing the residues gives the rational dependence on $\rho$ of Eq.~\eqref{eq:law}; the unit norm of $\Phi$ fixes the magnitude, and an overall packet phase does not affect $P=|A|^2$. The intermediate formulas and the propagation checks are in the Supplemental Material~\cite{SM}.

\emph{A common touching power on both sides.}---The construction applies to $\lambda(t)=c_\pm|t|^\nu$ with the same positive integer $\nu$ on both sides. The power sets the shape of the canonical packet; at fixed $\nu$ the two packets differ only in scale. With
$$
q=2\nu+1,\qquad \rho=\left(\frac{c_+}{c_-}\right)^{1/q},
$$
the particular Mellin solution is
$$
\begin{gathered}
\mathcal G_\nu(w)\propto\frac{\beta_\nu^{\,w}\,\Gamma(w)}{\displaystyle\prod_{k=1}^{\nu}\Gamma[(2w+2k-1)/q]},\\[2pt]
\beta_\nu=\left(\frac2q\right)^{2\nu/q}e^{-\mathrm{i}\pi/(2q)},
\end{gathered}
$$
and a constant periodic factor, as in the quadratic case, gives for the normalized overlap~\cite{SM}
\begin{equation}
P_\nu(\rho)=\frac{4\rho}{q}\prod_{l=-\nu}^{\nu}
\left[1+2\rho\cos\!\left(\frac{\pi l}{q}\right)+\rho^2\right]^{(-1)^{\nu+l+1}}.
\label{eq:lawnu}
\end{equation}
For $\nu=2$ this is Eq.~\eqref{eq:law}. For every positive integer $\nu$ the maximum lies at $\rho=1$ and is unique, so the reduction of recapture by unequal rates is not special to the quadratic touching. The selection of the packet remains conditional on the regularity assumption, and for $\nu=1$ one more step is open: the regularization that yields the Mellin recurrence for $\nu\ge2$ does not apply directly~\cite{SM}. Numerical propagation tests the formula independently. For the 19 tabulated combinations with $\nu\le12$ and $1/4\le\rho\le2$, the absolute differences in recapture probability are below $2\times10^{-9}$, within the numerical uncertainties of at most $10^{-8}$~\cite{SM}. These comparisons support the law; they do not establish the open analytic steps.

\emph{Different touching powers.}---If approach and return use different powers, $\nu_-$ and $\nu_+$, the canonical packets differ in shape as well as in scale. To compare with a protocol controlled by one rate parameter, take $c_\pm=v^{\nu_\pm}$ and $q_\pm=2\nu_\pm+1$, where $v>0$ slows both halves; then $k_\pm=c_\pm^{1/q_\pm}$ and
$$
\rho=\frac{k_+}{k_-}=v^{\,\nu_+/q_+-\nu_-/q_-},
$$
so the scale ratio changes with $v$ even though both halves share one rate parameter. For $(\nu_-,\nu_+)=(1,2)$, numerical overlaps of the inverse-Mellin packets reach a maximum recapture probability of about $0.478$ near $\rho=1.04$, against about $0.476$ from the factorized expression of Ref.~\cite{SokolovskiPons2016}; the $1$--$2\%$ variation reported there refers to the finite parameter ranges examined~\cite{SM}. For this pair of powers $\rho=v^{1/15}$, which varies slowly with $v$ but still tends to zero as $v\to0$, so weak variation over the tested ranges does not establish rate-independent recapture in the slow limit.
\end{document}